# PHASE ROTATION, COOLING AND ACCELERATION OF MUON BEAMS: A COMPARISON OF DIFFERENT APPROACHES

G. Franchetti, S. Gilardoni, P. Gruber, K. Hanke, H. Haseroth, E. B. Holzer, D. Küchler,
A. M. Lombardi, R. Scrivens, CERN, Geneva, Switzerland

*Abstract*

Experimental and theoretical activities are underway at CERN [1] with the aim of examining the feasibility of a very-high-flux neutrino source ($\approx 10^{21}$ neutrinos/year). In the present scheme, a high-power proton beam (some 4 MW) bombards a target where pions are produced. The pions are collected and decay to muons under controlled optical condition. The muons are cooled and accelerated to a final energy of 50 GeV before being injected into a decay ring where they decay under well-defined conditions of energy and emittance.

We present the most challenging parts of the whole scenario, the muon capture, the ionisation-cooling and the first stage of the muon acceleration. Different schemes, their performance and the technical challenges are compared.

## 1 INTRODUCTION

A next generation neutrino source (Neutrino Factory) should have the following advantages with respect to today's sources: a higher flux ($10^{21}$ neutrinos/year vs. the present $10^{11}$), a higher energy (50 GeV vs. the present 20 GeV for the CERN Neutrinos to Gran Sasso project) and the possibility to control the flavour of the neutrino beam. A beam with the aforementioned characteristics can be obtained by bombarding a target with high-energy protons, collecting the produced pions and allowing them to decay in a controlled environment before accelerating the muons to the required 50 GeV. The muons decaying in a storage ring, whose straight sections are pointed towards the detectors, will produce neutrinos with the desired characteristics. The design of the pion collection system and the first stage of the muon acceleration (front-end) are extremely challenging due to the large emittances involved.

## 2 FRONT-END DESIGN

To meet the requirements of a Neutrino Factory, $10^{20}$ to $10^{22}$ muons at 50 GeV, within a transverse emittance of 1.5 cm·rad and a longitudinal emittance of 0.06 eV·s must circulate in the decay ring. International collaborations, for engineering and radio-protection considerations, have standardised on the power on target of 4 MW.

In the CERN reference scheme, a proton driver [2] would be built to deliver 4 MW of 2.2 GeV protons in bursts of 3.3 μs repeated at 75 Hz, i.e. $10^{23}$ protons/year (notice that through this paper an operational year is assumed to consist of $10^7$ seconds).

From these figures it follows that the system downstream of the target should have a yield of at least 0.001 muon/proton for a minimum Neutrino Factory.

A typical phase space plot of the pions after production is shown in Fig. 1: the beam radius is 30 cm, the divergence 200 mrad and the kinetic energy ranges from 0.05 to 1 GeV. The peak of the production is around 0.1 GeV kinetic energy, and 75% of the pions have energies between 0.050 and 0.3 GeV. For the case presented in Fig. 1, 0.017 $\pi^+$ are produced for each 2.2 GeV proton on a thin (26 mm) mercury target. These data are linearly extrapolated to a 300 mm target.

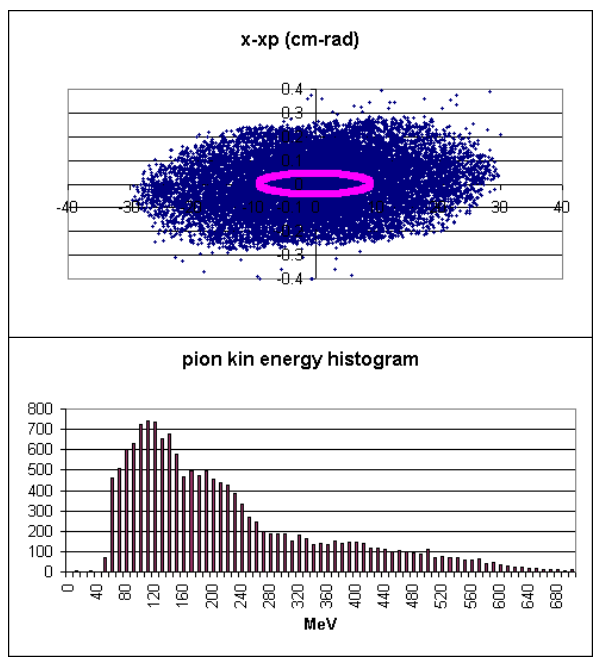

Figure 1: Transverse phase space and energy histogram of a pion beam produced by a 2.2 GeV proton beam impinging on a 26 mm Hg target immersed in a 60 cm bore 20 T solenoid (FLUKA calculations). The plots are at 4 m from the target. In the upper plot the acceptance of the decay ring is indicated for reference.

Insufficient pions fall within the energy-spread acceptance of any conventional accelerator. A reduction of energy spread by phase rotation can be envisaged in three possible ways:
- Apply it to the pions just after the target : this allows the full benefit of the defined time structure of the pions, but it implies the use of high-gradient rf cavities in a high radiation area. Further complications arise due to the decay of pions in an rf field.
- Allow the pions to decay (drift of tens of meters) before they enter an rf cavity with moderate gradients.
- Allow the pions to decay and let the muons drift for some hundreds of meters in order to build up a strong correlation between time and energy and then match the energy spread with a quasi-dc device (e.g. an induction linac).

The two latter approaches [3,4] will be described in detail in the next sections.

## 2.1 The rf approach to phase rotation

The parameters are presented in Table1.

The pions decay in a 30 m long channel and are focused by a 1.8 T solenoid. At the end of the decay channel, the beam enters a series of 44 MHz cavities and the energy spread of the particles with kinetic energy in the range 100-300 MeV is reduced by a factor two. A first cooling stage, employing the same rf cavities and 24 cm long $H_2$ absorbers, reduces the transverse emittance in each plane by a factor 1.4 while keeping the average energy constant. After the first cooling stage, the beam is accelerated to an average energy of 300 MeV. The beam phase extent, as well as the reduced physical dimensions of the beam, allows the continuation of the cooling with 88 MHz cavities. 40 cm long absorbers are used in this section. At the end of cooling, the emittance is reduced by a factor 4 in each transverse plane. The system is continued at 88 MHz, and 176 MHz until the energy of 2 GeV -suitable for injection in a Recirculating Linac Accelerator (RLA)- is reached. The system works also with 40 MHz, 80 MHz, 200 MHz.

## 2.2 The induction linac (IL) approach

The parameters are presented in Table 2.

The pions are transported through a tapered solenoid of initial field of 20 T, as favoured by the "Neutrino Factory and Muon Collider Collaboration" [5]. At 1.46 T the solenoid is continued for a distance of 200 m where the pion to muon decay occurs, and the strong momentum and time correlation is formed.

A high performance induction linac is then used to correct the energy spread of the beam. The induction linac is formed of 25 cm cells pulsed with a maximum voltage of 500 kV. The electrostatic field distribution was calculated using POISSON. A solenoid is placed inside each of the cells, producing an on-axis field of 1.46 T, rising to 1.9 T near the outer limits of the beam chamber, for which a large aperture of 60 cm is required. The linac produces an average gradient of 2 MV/m for a distance of 50 m. The correction affects the muons with kinetic energies between 120 and 310 MeV and results in a 330 ns long macro bunch. To obtain a bunched beam for cooling, a series of 176 MHz cavities are employed over a distance of 17 m followed by a further 17 m of drift space. The $\beta\lambda/2$ cavities are operated at 2 MV/m with 0.5 m solenoids between them, again providing an average on-axis field of 1.46 T. A cooling section follows in which the beam could be reduced in transverse emittance by a factor 3.

Table 1: RF solution, main parameters

|  |  | Decay | Rotation | Cooling I | Accel. | Cooling II | Accel |
|---|---|---|---|---|---|---|---|
| Length | [m] | 30 | 30 | 46 | 32 | 112 | ≈450 |
| Diameter | [cm] | 60 | 60 | 60 | 60 | 30 | 20 |
| B-field | [T] | 1.8 | 1.8 | 2.0 | 2.0 | 2.6 | 2.6 |
| Frequency | [MHz] | - | 44 | 44 | 44 | 88 | 88-176 |
| Gradient | [MV/m] |  | 2 | 2 | 2 | 4 | 4-10 |
| Kin Energy | [MeV] |  | 200 |  | 280 | 300 | 2000 |

Table 2: Induction Linac solution, main parameters

|  |  | Decay | Rotation | Bunching | Cooling |
|---|---|---|---|---|---|
| Length | [m] | 200 | 50 | 37 | 80 |
| Diameter | [cm] | 60 | 60 | 60 | 68 |
| B-field | [T] | 1.4 | 1.4 | 2.0 | 3 |
| Frequency | [MHz] | - | I.L. | 176 | 176 |
| Gradient | [MV/m] |  | ±2 | 2 | 15 |
| Kin Energy | [MeV] | - | 110 | 110 | 110 |

## 2.3 Particle budget

The particle budget for the two solutions studied is shown in Fig. 2. The number of muons delivered to the RLA is comparable for the two schemes and, assuming the CERN proton driver and target, the number of muons/year reaches $10^{21}$ in both.

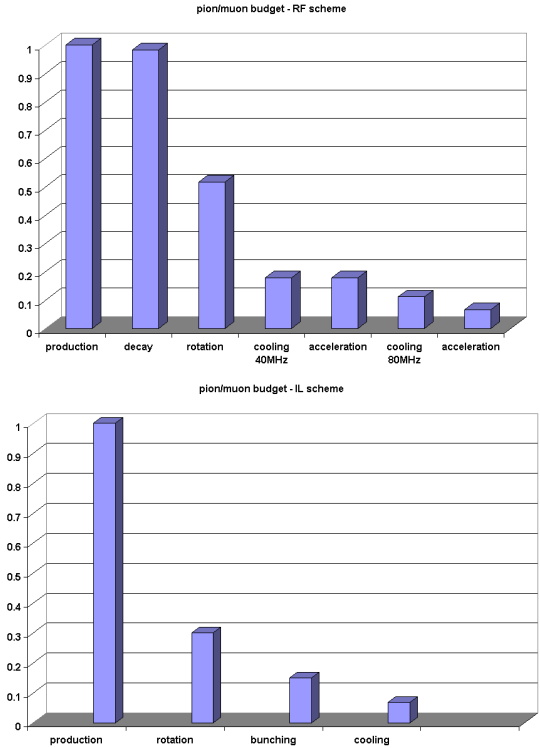

Figure 2: Particle budget along the rf and IL scheme.

In Fig. 2 we can identify the bottlenecks and the possible improvements to the designs. In the rf solution, the transition between phase rotation and the first stage of cooling should be smoothed in order to limit the losses (longitudinal losses, particles falling outside the bucket): interlacing of the two sections as well as the use of higher harmonics will be studied in the future. Besides, the result of rf tests should give a guideline for the maximum allowable gradient.

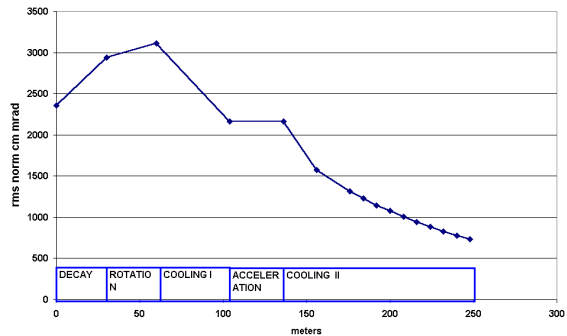

Figure 3: Emittance evolution in the rf solution

For the IL solution the improvement is needed in the bunching section before the cooling (50% efficiency at the moment).

In Fig. 3 the emittance evolution along the rf front end can be followed. An emittance increase by a factor 1.5 is generated during the decay. No effort has yet been dedicated to limiting this growth.

## 2.4 Comparison of the two approaches

Both the rf and IL approach deliver a sufficient number of neutrinos. The difference between them lies in the feasibility of the main components and on the constraints imposed on the proton driver. The Induction Linac parameters are very demanding and the limitation in the number of bunches/burst from the proton driver (12) would result in very high space-charge forces in the compressor ring. Conversely, the rf solution could accept any number of bunches from the accumulator (up to a max of 140). Preliminary calculations on a 44 MHz and a 88 MHz cavity show that an average power of 0.06 and 0.032 MW/cavity would suffice to provide the necessary field. The total average power (15 MW for the whole phase rotation and cooling rf system) is also considerably less than that needed for the IL alone (75 MW, scaling from [5]; although this can be reduced to 37 MW with a better core material).

## 3 CONCLUSIONS

Two possible schemes for the front-end of a neutrino factory have been explored at CERN. Both of them deliver the required number of neutrinos. The technical challenges, the CERN rf expertise [6] and the better match to the CERN proton driver have led to the choice of the rf scheme as the CERN reference scenario.

## REFERENCES

[1] H. Haseroth (for the NFWG), "Status of Studies for a Neutrino Factory at CERN", and R. Cappi et al., "Design of a 2 GeV Accumulator-Compressor for a Neutrino Factory", EPAC2000, Vienna.
[2] M. Vretenar, "A High-Intensity H⁻ Linac at CERN Based on LEP-2 cavities", these proceedings.
[3] A.M. Lombardi, "The 40-80 MHz Scheme", CERN-NUFACT-Note34.
[4] R. Scrivens, "Example Beam Dynamic Designs for a Neutrino Capture and Phase Rotation Line Using 50m, 100m and 200m Long Induction Linacs", CERN-NUFACT-Note14.
[5] N. Holtkamp, D. Finley (Eds.), "A Feasibility Study of a Neutrino Source Based on a Muon Storage Ring", FERMILAB pub-00/08-E, (2000).
[6] R. Garoby, D. Grier, E. Jensen, CERN. A. Mitra, R.L. Poirier, TRIUMF, "The PS 40 MHz Bunching Cavity", PAC'97, Vancouver, 1997.